\documentclass[amsmath,amssymb,nofootinbib]{revtex4}

\usepackage{bbold}
\usepackage{slashed}
\usepackage{bigints}
\usepackage{graphicx}

\begin{document}

\title{Completing the Theory of an Electron with Gravitational Torsion}

\author{Carl F. Diether, III}

\email{fred.diether@einstein-physics.org}
 
\author{Joy Christian}

\email{jjc@bu.edu}

\affiliation{Einstein Centre for Local-Realistic Physics, Oxford OX2 6LB, United Kingdom}

\maketitle

\begin{center}
\bf{Abstract}
\end{center}
\begingroup
\addtolength\leftmargini{0.2in}
\begin{quotation}
We demonstrate that classical and quantum  electrodynamics can be completed by gravitational torsion appearing in the Einstein-Cartan-Sciama-Kibble theory of gravity, providing the missing part of the theory of electron. One of the equations of this theory, a version of Dirac-type equation for fermions called the Hehl-Datta equation, contains a gravitational spin-torsion term in the Lagrangian density for quantum electrodynamics. This term relates the charged fermion spin to gravitational torsion and provides a mechanical energy counter-balance to the infinite electromagnetic self-energy. As a result, there is no ``bare" mass for an electron, nor is renormalization required for many scenarios.
This allows a completion of the theory of electron with gravitational torsion.
\end{quotation}
\endgroup

\parskip 5pt

\baselineskip 13.1pt

\section{Introduction}

\begin{flushright}
``You know, it would be sufficient to really understand the electron.''

--- Albert Einstein$\;\;$
\end{flushright}

Gravitational torsion may be viewed as a twisting of spacetime, analogous to the curving of spacetime accomplished by curvature tensor. In this regard, Einstein-Cartan-Sciama-Kibble (ECSK) theory of gravity turns out to be convenient \cite{Hehl1976, Trautman, Sciama, Kibble, Cabral, sabbata, Hehl2007, ohanian}, not the least because it has been thoroughly developed for over half a century. It is a minimal extension of general relativity that incorporates torsion and is in fact more general. For a recent review, see, for example, \cite{Fabbri}.  According to the ECSK theory, the spin of a fermion is a mechanism that twists spacetime. In this essay we argue that, in fact, a fermion would not exist without extreme twisting of spacetime. A fermion, in our view, is simply a twisted spacetime \cite{diether2}.

To that end, recall that if we solve the standard Lagrangian density of quantum electrodynamics (QED) for the radius of an electron, what we obtain is the classical electron radius, which can be deduced as follows. The standard Lagrangian density of QED is
\begin{equation}
\mathfrak{L}_{\rm QED}= i\hbar c\,\bar{\psi} \gamma^{\mu}\partial_{\mu}\psi + e\,\bar{\psi}\gamma^{\mu}A_{\mu}\psi - \frac{1}{4} F_{\mu\nu}F^{\mu\nu} - mc^2 \bar{\psi} \psi.
\end{equation}
If we set this equal to zero, then in the rest frame of the electron we obtain
\begin{equation}
 + e\,\bar{\psi}\gamma^{0}A_{0}\psi = mc^2 \bar{\psi} \psi\;\longrightarrow\;r = \frac{\alpha \hbar}{m_e c}.
\end{equation}
However, from scattering and other experiments \cite{Dehmelt} it is well known that the electron radius must be much smaller than this value. This suggests that QED of the Standard Model cannot be a complete theory of electrodynamics. On the other hand, if we use the Lagrangian density from which the Dirac-Hehl-Datta equation \cite{Hehl-Datta} is extracted that has a gravitational spin-torsion term in it and solve it for the electron radius, then we obtain two solutions for the radius: the classical radius and another  close to Plank length. This raises the question, why would an electron have two radial ``sizes"?  We will try to answer that question in this essay.

\section{The Gravitational Spin-Torsion Term}

In 1971, Hehl and Datta formulated a Dirac type of equation for fermions \cite{Hehl-Datta} with a spin-torsion term derived within the ECSK theory of gravity. We will call this equation the Hehl-Data equation. It has the following form:
\begin{equation}
i\gamma^{\alpha} \nabla_{:\alpha}\psi - \frac{3}{8} l_P^2(\bar{\psi}\gamma_5 \gamma^{\alpha}\psi)\gamma_5 \gamma_{\alpha}\psi = m\psi\,,
\end{equation}
where the colon denotes the covariant derivative and $l_P$ is Plank length. The second term on the left hand side is the gravitational spin-torsion term. If we include that term in the Lagrangian for quantum electrodynamics \cite{Perez, Freidel, Freidel-2}, then we have
\begin{equation}
\frac{\mathfrak{L}_{\rm QED}}{\sqrt{-g\,}}= i\hbar c\,\bar{\psi} \gamma^{\mu}\partial_{\mu}\psi + e\,\bar{\psi}\gamma^{\mu}A_{\mu}\psi - \frac{1}{4} F_{\mu\nu}F^{\mu\nu} - mc^2 \bar{\psi} \psi - \frac{3\kappa \hbar^2 c^2}{16}(\bar{\psi}\gamma^5\gamma_{\mu}\psi)(\bar{\psi}\gamma^5\gamma^{\mu}\psi).\label{rf1}
\end{equation}
We believe this Lagrangian to be the nearly complete QED Lagrangian for charged fermions. It is a generalization of the Lagrangian in eq.~(1). It can be viewed as a postulate or hypothesis that allows a point-like particle (elementary fermion) to have a small rest mass-energy instead of infinite or huge rest mass-energy. It is a reasonable hypothesis because it is not contrary to experimental evidence.  However, eq.~(4) is not a complete generalization, since it does not include normal (or curvaturial) gravity. Since for a single electron normal gravity is essentially zero, it is reasonable to keep only torsional gravity in the Lagrangian.

If we now move to the rest frame of a charged elementary fermion, then we will be concerned only with the following terms of the Lagrangian:
\begin{equation}
\mathfrak{L}_{\rm QED,R.F.} = + e\,\bar{\psi}\gamma^{0}A_{0}\psi - mc^2 \bar{\psi} \psi - \frac{3\kappa \hbar^2 c^2}{16}(\bar{\psi}\gamma^5\gamma_{0}\psi)(\bar{\psi}\gamma^5\gamma^{0}\psi). \label{rf}
\end{equation}
We have used this form in our derivation in the Appendix. The usual gravitational effects undoubtedly vanish in the rest frame of a single elementary fermion such as an electron, but the gravitational spin-torsion term does not vanish at lengths smaller than the classical radius computed above. It is easy to demonstrate this for an electron as follows.

\begin{figure}[t]
\vspace{-0.3cm}
\centering
\includegraphics[scale=0.6]{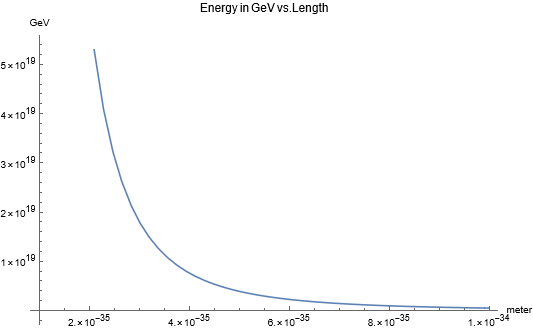}
\caption{Energy of the spin-torsion term near Planck length.}
\vspace{0.2cm}
\label{Fig-1}
\end{figure}

\begin{figure}[t]
\vspace{-0.3cm}
\centering
\includegraphics[scale=0.6]{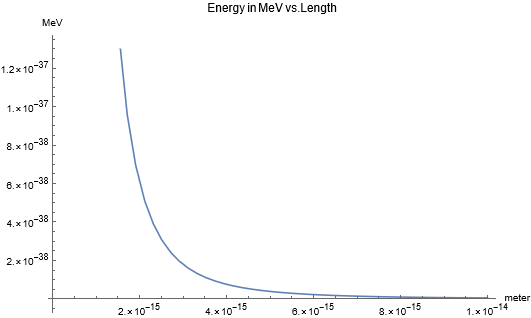}
\caption{Energy of the spin-torsion term near the classical radius.}
\vspace{0.2cm}
\label{Fig-2}
\end{figure}

By setting the Lagrangian in Eq.~(\ref{rf}) equal to zero, it can be solved for a function of the radius, as shown in our semi-classical derivation in the Appendix.  The result for an electron is,
\begin{equation}
\frac{\alpha \hbar c}{r} - \frac{3\kappa (\hbar c)^2}{8\,r^3} = m_e c^2,  \label{electron}
\end{equation}
where $r$ is the radius, $\kappa = 8\pi G/c^4$, and $\alpha = e^2/4\pi\hbar c$ is the fine structure constant.  It is easy to see that the first electrostatic term is balanced by the second term to produce the rest mass-energy of the electron.  However, if we substitute for the classical radius, $r = \alpha\hbar c/m_e c^2$, in this equation, then we obtain the following value for the spin-torsion term,
\begin{equation}
  \frac{3\kappa (\hbar c)^2}{8\,(\frac{\alpha\hbar c}{m_e c^2})^3} = \frac{3\pi G \hbar^2}{c^2 (2.818\times 10^{-15}\,\text{meter})^3} \simeq 2.171 \times 10^{-38}\,\text{MeV}.  \label{electron2}
\end{equation}
Thus, near the classical radius the spin-torsion term produces negligible amount of energy. Now, if we set the radius to say, $r = 10^{-22}\,\text{meter}$ for which it is thought that an electron must have a radius smaller than that, then we obtain
\begin{equation}
  \frac{3\pi G \hbar^2}{c^2 (10^{-22}\,\text{meter})^3} \simeq 4.858 \times 10^{-16}\,\text{MeV},  \label{electron3}
\end{equation}
which is still a rather small amount of energy.  And finally, if we use Planck length for the radius, then we obtain
\begin{equation}
  \frac{3\pi G \hbar^2}{c^2 \Big(\sqrt{\frac{G \hbar}{c^3}}\Big)^3} \simeq 1.151 \times 10^{20}\,\text{GeV},  \label{electron4}
\end{equation}
which is a huge amount of energy for an elementary electron. However, it is negative energy relative to the field produced by the electrostatic term.  The important thing to realize here is that the field produced by the spin-torsion term balances out the electrostatic energy and provides a natural cut-off near Planck length so that the self-energy of an electron is not infinite.  The radius near Planck length works out to be $\simeq 5.808 \times 10^{-34}$ meter. It is remarkable that we still obtain the normal rest mass-energy for the electron with this very small radial ``size." From the plots presented in Fig.~\ref{Fig-1} and Fig.~\ref{Fig-2} we can see how the spin-torsion term varies as a function of length. From Fig.~\ref{Fig-2} we see that the value of the spin-torsion term is negligible compared to that in Eq.~(\ref{electron3}) as the length grows larger. Perhaps this is the reason why it has not been experimentally detected.

For comparison, the energy of the electrostatic term at Planck length is
\begin{equation}
\frac{\alpha \hbar c}{\Big(\sqrt{\frac{G \hbar}{c^3}}\Big)} \simeq \,8.909\times 10^{16}\,\text{GeV}.  \label{electron5}
\end{equation}
However, it turns out that this does not balance out the spin-torsion term energy despite the fact that very near to the $5.808 \times 10^{-34}$ meter solution the electrostatic term and spin-torsion term appear to be equal to each other.  In fact, that solution is about 36 times longer than Planck length.  When the two terms are equal, for $r_t$ we have
\begin{equation}
r_t = \sqrt{{\frac{3\pi}{\alpha\,}}}\;l_P\,\simeq \,  5.80838808109165274355010 \times 10^{-34}\; \text{meter},
\end{equation}
using arbitrary precision in Mathematica \cite{Diether}. Note that $r_t$ is entirely composed of known constants. For comparison, the solution for an electron is
\begin{equation}
    r_e \simeq 5.80838808109165274414872 \times 10^{-34}\; \text{meter}.
\end{equation}
Thus, it is very sensitive at this length to 19 significant figures.

\section{Comparison with Standard Electron Theory}

In standard electron theory, the observed mass is supposedly composed of the electromagnetic mass, $\delta m$, plus the bare mass, $m_0$, which would have to be a negative mechanical mass of some kind.  We believe we have found this negative mechanical mass via the gravitational spin-torsion mechanism. In 1939, Weisskopf calculated the electromagnetic mass and found it to be logarithmically divergent \cite{Weisskopf}.  Ever since then it has been thought that the mechanical ``bare" mass would have to be negatively divergent.  But with gravitational spin-torsion added to the scenario makes that possibility incorrect.  It turns out that neither the electromagnetic mass nor the ``bare" mass are divergent because they are automatically cancelling each other out, leaving only some electromagnetic mass for the rest mass.  In other words, it turns out that the rest mass of an electron is basically entirely electromagnetic, as can be seen in our Eq.~(\ref{electron}).

After a few attempts using standard electron theory with the spin-torsion term, we realized it was completely the wrong process to use for electron self-energy.  The proper process is to analyze in the rest frame only of the charged fermion.  This became clear from our semi-classical analysis which can be found in the Appendix of this paper.  The implication is that there is no linear momentum at all and there are no propagators to use from Feynman diagrams.

 Upon first appearances, for the rest frame, it may seem that the middle term in the non-linear spin-torsion term, $(\bar{\psi}\gamma_k\gamma5\psi)$, vanishes because $(\bar{\psi}\gamma_0\gamma5\psi) = 0$ normally, which makes the whole spin-torsion term vanish in the rest frame. By contrast, we have found that fermion anti-fermion mixing is involved in the spin-torsion term in the rest frame of the particle, as we demonstrate below in the Appendix. Moreover, in the full machinery of quantum field theory, $\psi$ can represent a particle or anti-particle in the rest frame, as in the following equations:
\begin{equation}
\psi(0, t)\mid i\,\rangle = \sqrt{\frac{1}{r^3}} \,u^i(m)e^{-iEt} \quad \text{or} \quad \psi(0, t)\mid i\,\rangle = \sqrt{\frac{1}{r^3}} \,v^i(m)e^{+iEt},
\end{equation}
where the $\mid i\,\rangle$ represents the initial state with $u^i$ for the particle and $v^i$ the anti-particle. In order for the spin-torsion term to be non-zero in the rest frame, one must have one of the $\psi$'s represent an appropriate anti-fermion, as in the following equations: 
\begin{equation}
\left(\begin{array}{cccc}
1& 0& 0& 0
\end{array}\right)
\gamma^5 \left(\begin{array}{c}
0\\
0\\
1\\
0 
\end{array}\right) = 1\quad \text{or} \quad
\left(\begin{array}{cccc}
0& 0& 1& 0
\end{array}\right)
\gamma^5 \left(\begin{array}{c}
1 \\
0 \\
0 \\
0 
\end{array}\right) = 1.
\end{equation}
This fermion anti-fermion mixing is somewhat of a mystery here, but highly probable by interaction with the quantum vacuum.  However, the mixing is certainly necessary, otherwise the electron would blow itself up with infinite energy if the spin-torsion term were zero.

\begin{figure}[t]
\vspace{-0.3cm}
\centering
\includegraphics[scale=0.6]{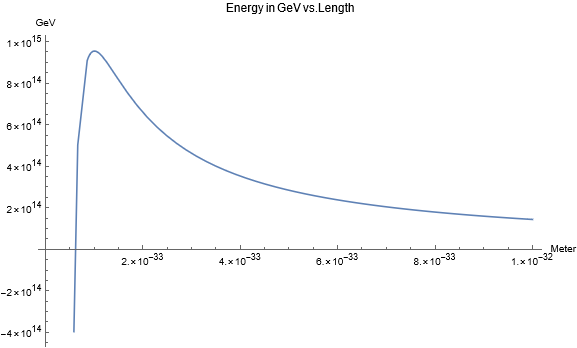}
\caption{$E(r) = {\alpha \hbar c}/{r} - {3\kappa (\hbar c)^2}/{8\,r^3}$ near Planck length.}
\vspace{0.2cm}
\label{Fig-3}
\end{figure}

\section{Further Discussion and Open Questions}

We considered a Planck ``sized" electron that is an electro-weak particle as far as its energy is concerned.  The question now is, why does it appear to have two different radii? Why does it have the rest mass-energy at the classical radius and also at the radius near Planck length? Fig.~\ref{Fig-3} shows the behavior of the function, $E(r)$, of the electrostatic energy of an electron minus the spin-torsion energy as a function of length that explains a part of this question:  
\begin{equation}
E(r) = \frac{\alpha \hbar c}{r} - \frac{3\kappa (\hbar c)^2}{8\,r^3}.  \label{energy1}
\end{equation}
The solution for $m_e c^2$ near Planck length is on the very steep slope on the left hand side of the plot whereas the solution at the classical radius is way off to the right hand side of the figure.  Notice also the peak energy of about $9.54173\times 10^{14}$ GeV at length of about $1.006\times 10^{-33}$ meters.  This is reminiscent of GUT scale energies. Could there be charged fermions with this much energy?  And what constrains the electron's energy to be $m_e c^2$?  These are some questions that remain to be investigated.  This scenario also applies to muons and tauons, even though they are not stable particles.  For quarks, $E(r)$ becomes much more complicated because in that case color charge is added to the electromagnetic charge.  For neutrinos, the charge would be the weak charge in formula for $E(r)$, requiring a replacement of $\alpha$ in it for electromagnetics with $\alpha_w$ for weak charge.

We suspect that the classical radius is primarily due to screening of the charge by virtual particles from the quantum vacuum.  If it were possible to see near Planck length via scattering of some kind after breaking through the screening effect, one would see the radius near Planck length.  The screening effect of interaction with the quantum vacuum ``spreads" out the apparent charge of an electron so that one might think there is a ``bare" charge if the screening was not there \cite{Milonni}. This is suggested by Schrodinger's conception of Zitterbewegung. Accordingly, the point-like entity of the electron near Planck length vibrates around in the quantum vacuum interacting with virtual electrons and positrons with an amplitude of about the classical radius.

It is instructive here to investigate the ``standard" result for the logarithmically divergent electron self-energy, which is the orthodox solution for electromagnetic mass, $\delta m$ \cite{Milonni, Weisskopf}, so that
\begin{equation}
    \delta m  = \frac{3 m \alpha}{2\pi}\log\frac{\Lambda}{m}, \label{first}
\end{equation}
where $\Lambda$ is an arbitrary cutoff that, when taken to infinity, gives the logarithmically divergent result. However, the $m$ in this expression enters via a propagator when using a quantum field theory S-matrix evaluation. It is thus virtual and can be of any value. That allows us to make a substitution $m = 1/r$ so that,
\begin{equation}
    \delta m  = \frac{3 \alpha}{2\pi r}\log\frac{r}{\Lambda},
\end{equation}
which is an expression similar to that for electrostatic energy, and $\Lambda$ is now a length instead mass. This expression is still logarithmically divergent as $\Lambda \rightarrow 0$ or $r\rightarrow 0$.  If we now set $r = \Lambda$, we get zero for electromagnetic mass. Thus the expression is quite meaningless.  In fact, for $m = \Lambda$ in Eq.~(\ref{first}) we also get zero, which is possible since the mass is virtual. That is the main reason why our attempts to use the ``standard" result failed. As shown by Milonni in \cite{Milonni}, the observed mass is in fact
\begin{equation}
    m_{obs} = m_0 + \delta m = m_0 + \frac{3 m_0 \alpha}{2\pi}\log\frac{\Lambda}{m_0},
\end{equation}
where $m_0$ is the ``bare" mass. Since $m_0$ is in the $\delta m$ part of the result, that reinforces our notion of setting $m=1/r$. Moreover, the observed mass is dependent on an arbitrary cutoff.  The argument to fix that is that the ``bare" mass also depends on the cutoff.  That seems fairly convoluted.  By contrast, our strategy is simpler. All that is needed is subtraction of the ``bare" mass (spin-torsion term) from the electrostatic mass to obtain the observed mass in the rest frame. It is remarkable that the fairly simple spin-torsion term completes electrodynamics in this manner. No renormalization procedure is required for many calculations. But it could still be useful for some.

\section*{Acknowledgement}
The authors would like to thank Luca Fabbri for encouragement and discussions about gravitational torsion.

\appendix*
\section{Semi-Classical Derivation of Charged Fermion Self-Energy}

In the rest frame, where normal gravity is effectively zero for an elementary fermion, with natural units $\hbar = c = 1$, along with the electrostatic term via a local gauge transformation, we have from the Lagrangians (\ref{rf1}) and (\ref{rf}),
\begin{equation}
i \,\gamma^0 \frac{\partial\psi}{\partial t\,} + q A_0\,\gamma^0\psi = m \, \psi + \frac{3\kappa}{8} \left(\bar{\psi}\gamma^5 \gamma_0\psi\right) \gamma^5\gamma^0\psi\,, \label{restbc}
\end{equation}
which can be further simplified to
\begin{equation}
i\left(\begin{array}{c}
+\frac{\partial\psi_1}{\partial t\;} \\ \\
+\frac{\partial\psi_2}{\partial t\;} \\ \\
-\frac{\partial \psi_3}{\partial t\;} \\ \\
-\frac{\partial \psi_4}{\partial t\;} 
\end{array}\right)
+ q A_0
\left(\begin{array}{c}
+\psi_1 \\ \\
+\psi_2 \\ \\
-\psi_3 \\ \\
-\psi_4
\end{array}\right)
= m
\left(\begin{array}{c}
+\psi_1 \\ \\
+\psi_2 \\ \\
+\psi_3 \\ \\
+\psi_4
\end{array}\right)
- \frac{3\kappa}{8}\left\{\psi_1^*\psi_3+\psi_2^*\psi_4+\psi_1\psi_3^*+\psi_2\psi_4^* \right\}
\left(\begin{array}{c}
-\psi_3 \\ \\
-\psi_4 \\ \\
+\psi_1 \\ \\
+\psi_2
\end{array}\right), \label{A2}
\end{equation}
where we have used 
\begin{equation}
\gamma^0 = \left(\begin{array}{cccc}
+1& 0& 0& 0\\
0& +1& 0& 0\\
0& 0& -1& 0\\
0& 0& 0& -1
\end{array}\right)\;\;\;\;\;\;\;\;\;\;\text{and}\;\;\;\;\;\;\;\;
\gamma^5= \left(\begin{array}{cccc}
0& 0& +1& 0 \\
0& 0& 0& +1 \\
+1& 0& 0& 0 \\
0& +1& 0& 0
\end{array}\right).
\end{equation}
If we now represent the particles and anti-particles with two-component spinors ${\psi_a}$ and ${\psi_b}$, respectively \cite{Griffiths}, where
\begin{equation}
\psi_a := \left(\begin{array}{c}
\psi_1 \\
\psi_2 \end{array}\right)
\;\;\;\text{and}\;\;\;
\psi_b := \left(\begin{array}{c}
\psi_3 \\
\psi_4 \end{array}\right) \label{c444}
\end{equation}
are the two-component spinors constituting the four-component Dirac spinor, then the above equation (\ref{restbc}) can be written as two {\it coupled} partial differential equations:
\begin{align}
+\,i \,\frac{\partial \psi_a}{\partial t\;} + q A_0 \,\psi_a &= m\,\psi_a + \frac{3\kappa}{8} \left\{\psi_1^*\psi_3+\psi_2^*\psi_4+\psi_1\psi_3^*+\psi_2\psi_4^*\right\}\,\psi_b \label{10bc}\\
-\,i \,\frac{\partial \psi_b}{\partial t\;} - q A_0 \,\psi_b &= m\,\psi_b - \frac{3\kappa}{8} \left\{\psi_1^*\psi_3+\psi_2^*\psi_4+\psi_1\psi_3^*+\psi_2\psi_4^*\right\}\,\psi_a\,. \label{11bc}
\end{align}
It is now very easy to see the fermion anti-fermion mixing in the spin-torsion non-linear term, which can be seen also in Eq.~(\ref{A2}).

Unlike the case in Dirac equation, these equations for the spinors ${\psi_a}$ and ${\psi_b}$ are coupled equations even in the rest frame. They decouple in the limit when the torsion-induced axial-axial self-interaction is negligible. On the other hand, at low energies it is reasonable to assume that, in analogy with the Dirac spinors in flat spacetime, the above two-component spinors for free particles decouple in the rest frame, admitting plane wave solutions of the form
\begin{equation}
\psi_a(t) = \sqrt{\frac{1}{V}}\,e^{-iEt} \psi_a(0) \;\;\;\;\;\text{and}\;\;\;\;\;
\psi_b(t) = \sqrt{\frac{1}{V}}\,e^{+iEt} \psi_b(0), \label{psiabc}
\end{equation}
where $V = r^3$ is spatial volume and $E=m=\omega = 2\pi/t$ with $\hbar = 1$ is frequency in the rest frame. The rest frame is a slice in time ({\it i.e.}, a space-like hypersurface), and therefore the time-derivative within it is zero. Consequently, in the rest frame the derivative term in the Eqs.~(\ref{10bc}) and (\ref{11bc}) vanishes since $\psi$ is constant and the kinetic energy in the rest frame would be zero, and also because in the rest frame the probability of finding the particle in a given volume at time ${t}$ is one. Moreover, $E = 2\pi/t$ = angular frequency $\omega$, gives
\begin{equation}
\psi_a(t) = \psi_b(t). \label{14bc}
\end{equation}
Consequently Eqs.~(\ref{10bc}) and (\ref{11bc}) uncouple and simplify to 
\begin{align}
+ q A_0 \,\psi_a &= m\,\psi_a + \frac{3\kappa}{8} \left\{\psi_1^*\psi_1+\psi_2^*\psi_2+\psi_3\psi_3^*+\psi_4\psi_4^*\right\}\,\psi_a\,, \label{10bcu}\\
- q A_0 \,\psi_b &= m\,\psi_b - \frac{3\kappa}{8} \left\{\psi_1^*\psi_1+\psi_2^*\psi_2+\psi_3\psi_3^*+\psi_4\psi_4^*\right\}\,\psi_b\,. \label{11bcu}
\end{align}
Substituting the $\psi$'s from Eq.~(\ref{psiabc}) and simplifying reduces Eqs.~(\ref{10bcu}) and (\ref{11bcu}) to the following pair of equations:
\begin{align}
+ q A_0 \psi_a(0) -  \frac{3\kappa}{8\,r^3}\,|\psi(0)|^2 \,\psi_a(0) &= m\,\psi_a(0) \label{16abc} \\
\text{and}\;\;\;
- q A_0 \psi_b(0) +  \frac{3\kappa}{8\,r^3}\,|\psi(0)|^2 \,\psi_b(0) &= m\,\psi_b(0) \,. \label{16bbc}
\end{align}
And since $|\psi(0)|^2 = 1$ in the rest frame, these equations can be further simplified to
\begin{align}
+\,q A_0 - \frac{3\kappa}{8\,r^3}&= m \,,  \\
\text{and}\;\;\;
-\,q A_0 + \frac{3\kappa}{8\,r^3}&= m \,.
\end{align}
Substituting in natural units for the scalar field ${A_0= V = q /(4\pi r)}$ in the Lorentz gauge (where ${V}$ is the electric potential), and for $\kappa= 8\pi G$, we finally arrive at our central equations, for any electroweak fermion of charge ${q}$ and mass ${m}$ and its anti-particle in the Riemann-Cartan spacetime:
\begin{align}
+\frac{q^2}{4\pi r} - \frac{3\pi G}{r^3} &= +m  \label{resultbc} \\
\text{and}\;\;\;\;
-\frac{q^2}{4\pi r} + \frac{3\pi G}{r^3} &= +m, \label{anti-resultbc}
\end{align}
where ${r}$ is the radial distance from ${q}$ and the two equations correspond to the particle and anti-particle, respectively.  Finally, replacing $q$ with electron or positron charge and replacing $\hbar$ and $c$ gives us our central equation for a fermion for our semi-classical evaluation:
\begin{equation}
\frac{\alpha \hbar c}{r} - \frac{3\kappa (\hbar c)^2}{8\,r^3} = m c^2.  \label{result2bc} 
\end{equation}
This same equation is also derived via a S-matrix quantum field theory evaluation for the rest frame in \cite{Diether}.

\end{document}